# Two-dimensional electrons occupying multiple valleys in AlAs

M. Shayegan,[*] E.P. De Poortere, O. Gunawan, Y.P. Shkolnikov, E. Tutuc, and K. Vakili

Department of Electrical Engineering, Princeton University, Princeton, NJ 08544, USA

Two-dimensional electrons in AlAs quantum wells occupy multiple conduction-band minima at the X-points of the Brillouin zone. These valleys have large effective mass and g-factor compared to the standard GaAs electrons, and are also highly anisotropic. With proper choice of well width and by applying symmetry-breaking strain in the plane, one can control the occupation of different valleys thus rendering a system with tuneable effective mass, g-factor, Fermi contour anisotropy, and valley degeneracy. Here we review some of the rich physics that this system has allowed us to explore.

## 1  Introduction

In selectively-doped semiconductor structures the carriers are spatially separated from the dopant atoms to reduce the ionized impurity scattering. As a result, the low-temperature mobility of the carriers in these structures far exceeds the values typically observed in uniformly-doped materials.[1] Thanks to the reduced disorder, these structures provide nearly ideal systems for two of the most exciting areas of solid state physics and device research. One is the ballistic and quantum (wave) transport phenomena whose study is made possible by the long carrier mean-free-path and phase coherence length in such materials. With the help of high-resolution lithographic and novel growth techniques to produce low-dimensional structures, a wealth of new, mesoscopic phenomena and novel device concepts have emerged in recent years. The other area of research afforded by these materials deals with the fascinating and often unexpected collective states arising from the strong electron-electron interaction.

In the past, much of the work on such clean systems has focused on the two-dimensional electron system (2DES) confined to a GaAs layer in a selectively-doped GaAs/AlGaAs heterojunction. Although other materials, such as Si/SiGe heterostructures, have certainly provided exciting results, GaAs/AlGaAs remains the system of choice for exploring many of the fundamental physical properties of electrons in low-dimensional systems. This popularity arises, on the one hand, from the nearly ideal combination of two closely lattice-matched crystals, GaAs and AlAs and, on the other hand, from the cleanliness of the molecular beam epitaxy (MBE) systems that are used to grow these structures. Indeed, the 2DES confined to a GaAs layer provides the cleanest system yet available as measured by, e.g., its low-temperature mobility. In the vast majority of GaAs/AlGaAs structures the electrons are confined in GaAs while AlAs, or more commonly AlGaAs, is used as the barrier material. In such structures the electrons occupy the GaAs conduction band minimum at the Γ-point of the Brillouin zone (Fig. 1a). It turns out, however, that in a properly designed structure, containing a pure AlAs layer and selectively-doped AlGaAs barriers, one can confine electrons to the AlAs layer.[2-9] In this case, the electrons occupy the AlAs X-point conduction band valleys (Fig. 1b) and have properties that are quite distinct from GaAs 2DESs. Here we review the basic properties of the 2DES confined to an AlAs quntum well (QW), and some of the novel phenomena that we have observed in this system.

[*] Corresponding author: e-mail: shayegan@princeton.edu, Phone: +001 609 258 4639, Fax: +001 609 258 6279

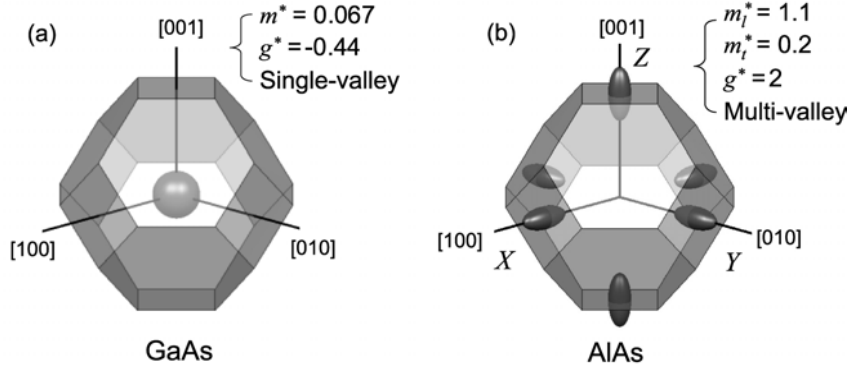

**Fig. 1** Schematic drawing of the Brillouin zone and constant energy surfaces of the lowest energy bands for (a) bulk GaAs and (b) bulk AlAs.

## 2   Realization of high-mobility 2D electrons in AlAs quantum wells

### 2.1 Band structure and valley occupation in AlAs quantum wells

Bulk AlAs has energy band minima at the Brillouin zone X-points, giving rise to ellipsoidal conduction electron Fermi surfaces (Fig. 1b). The electrons have a large and anisotropic effective mass (longitudinal $m_l^* = 1.1 \pm 0.1$, transverse $m_t^* = 0.20 \pm 0.02$) in contrast to the much lighter and isotropic mass ($m^* = 0.067$) of electrons in GaAs (all effective masses are given in units of the free electron mass).[10] The effective Lande g-factor of electrons in bulk AlAs ($g^* = 2$) is also much larger in magnitude and of a different sign than in GaAs ($g^* = -0.44$). Moreover, the electrons occupy multiple conduction band valleys in AlAs. These three main characteristics also differentiate 2DESs in modulation-doped AlAs QWs from those in GaAs QWs and lead to novel phenomena as we discuss in this paper.

Note that the band structure of AlAs electrons is similar to Si, except that in Si there are six ellipsoids centered around six equivalent points along the Δ-lines of the Brillouin zone, while in AlAs we have *three* (six half) ellipsoids at the X-points. We denote these valleys by the directions of their major axes: *X*, *Y*, and *Z* for the [100], [010], and [001] valleys, respectively. Another important difference between AlAs and Si is the manner in which the valleys are occupied in a QW. When electrons are confined along the [001] direction in a (001) Si-MOSFET or a Si/SiGe heterostructure, the two *Z* valleys, with their major axes pointing out of plane, are occupied because the larger mass of electrons along the confinement direction lowers their energy. In an AlAs QW grown on a (001) GaAs substrate, however, the *Z* valley is occupied only if the well thickness is less than approximately 5 nm (note that the growth direction is along *z*, *i.e.*, [001]). For larger well thicknesses, a biaxial compression of the AlAs layer, induced by the lattice mismatch between AlAs and GaAs, causes the *X* and *Y* valleys with their major axes lying in the plane to be occupied.[2-9]

### 2.2 Fabrication of high-mobility 2D electrons in AlAs quantum wells

During the past few years, in our laboratory we have concentrated on improving the quality of 2DESs confined to AlAs QWs. The basic sample structure and conduction-band alignment are shown in Fig. 2. Starting with an undoped GaAs substarate, we use MBE to grow a layer of AlAs flanked by layers of $Al_xGa_{1-x}As$ with x ≈ 0.4. The top AlGaAs layer is doped with Si at a distance of 75 nm away from the AlAs layer and the structure is finished with a cap layer of GaAs. We make our samples typically in a Hall bar geometry, fit them with back- and front-side gate electrodes, and use alloyed AuGeNi for ohmic contacts. The samples exhibit a pronounced *field-effect* persistent photo-conductivity,[11] which we utilize to populate the AlAs QW with electrons.

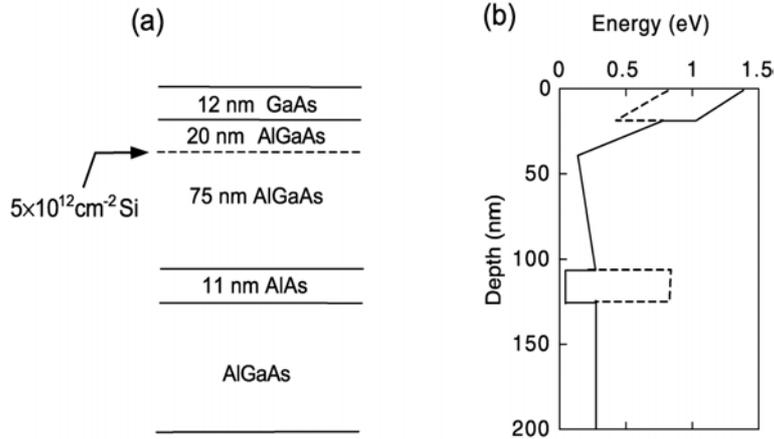

**Fig. 2** (a) Structure of a typical, selectively-doped AlAs quantum well. (b) X-point and Γ-point conduction band edges are indicated by solid and dashed lines, respectively. For $Al_{0.4}Ga_{0.6}As$, which is used as the barrier material, these energies nearly coincide. [After Ref. 7.]

We have made breakthroughs by producing both wide[7] *and* narrow[9] AlAs QWs with record-high low-temperature mobilities. In wide QWs, where the in-plane valleys (*X* and *Y*) are occupied, the mobilities of our best samples exceed $3 \times 10^5$ $cm^2/Vs$.[7,12] These samples exhibit a host of phenomena including ballistic transport, giant piezo-resistivity, and many-body effects such as quantum Hall effect (QHE) at integer and fractional filling factors and ferromagnetism involving valley and/or spin degrees of freedom; we will discuss some examples of these here. We have also succeeded in fabricating 2DESs confined to narrow (< 5 nm-wide) AlAs QWs where the electrons occupy the *Z* valley.[9] The low-temperature mobility in this case is limited by interface roughness scattering;[13] nevertheless, we have achieved the highest mobilities yet reported, $\sim 5 \times 10^4$ $cm^2/Vs$.[9,13] This is more than an order of magnitude larger than the highest previously reported mobility values for similar types of samples[6] and comparable to the highest mobilities reported in Si-MOSFET 2DESs.[14] And again, the samples reveal novel behavior, e.g., an interaction-induced enhancement of the spin-susceptibility at low densities that is the largest among all 2DESs and is remarkably close to that of an ideal, interacting 2DES.[9]

### 2.3 Tuning the valley population via application of in-plane strain, and giant piezo-resistivity

One of our major, recent accomplishments was our ability to change and monitor the occupancy of the *X* and *Y* valleys in our wide AlAs QWs. We developed a simple technique to induce symmetry-breaking strain in the sample plane at low temperatures and change the populations of the *X* and *Y* valleys.[15] We simply glue a thinned sample, made in the shape of a Hall bar, on the side of a commercial, stacked, piezoelectric actuator (Fig. 3a). The piezo's polling direction is aligned along the sample's [100] crystal orientation. When bias ($V_P$) is applied to the piezo stack, it expands (shrinks) along the [100] direction for $V_P > 0$ ($V_P < 0$) and shrinks (expands) in the [010] direction. We have confirmed that this deformation is fully transmitted to the sample at low temperatures and, using metal strain gauges glued on the opposite side of the piezo (Fig. 3a), have measured its magnitude in both [100] and [010] directions.[15] Such strain results in a splitting between the *X* and *Y* valley energies that is equal to, in the absence of interaction, $\varepsilon E_2$, where $\varepsilon = (\Delta L/L)_{[100]} - (\Delta L/L)_{[010]}$, and $E_2 = 5.8 eV$ is the AlAs deformation potential for the splitting between the *X* and *Y* valley energies ($\Delta L/L$ is the fractional change in sample size).[16] Figure 3b provides exemplary data of the sample resistance as a function of $V_P$.[17]

The piezo-resistance trace of Fig. 3b can be understood qualitatively based on the occupation of the in-plane valleys. In the absence of in-plane strain, the *X* and *Y* valleys are equally occupied. As the sample is compressed along the [100] direction and dialated along [010], the energy of the [100] valley is lowered with respect to the [010] valley and there is a charge transfer from *Y* to *X* while the total density

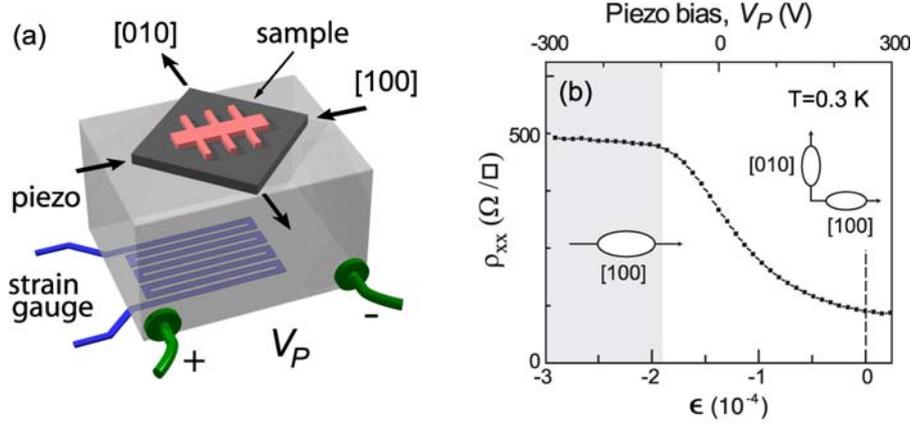

**Fig. 3** (a) Setup for inducing in-plane stress in a sample containing an AlAs 2DES. (b) Piezoresistivity of a 2DES confined to an 11 nm-wide AlAs QW. The [100] and [010] valleys are equally occupied at zero strain, while for a sufficiently large compression of the sample along [100], all the electrons are transferred to the [100] valley (shaded region). Note that in this experiment the sample was under compression at zero piezo bias because of the difference in its contraction with respect to the piezo during the cooling to low temperatures. [After Refs. 15 and 17.]

remains fixed. For sufficiently large compressive strain, all the electrons are transferred to the $X$ valley. Since for this valley the transport effective mass for electron motion along [100] ($m_l^*$) is larger than along [010] ($m_t^*$), the resistivity along [100] also becomes larger (Fig. 3b). Note that in our experiments, we can deduce the valley populations directly from the frequencies of the Shubnikov – de Haas oscillations.[18,19] Also, we would like to emphasize that the data of Fig. 3b represent a giant piezo-resistivity effect: near $\varepsilon = -1.4 \times 10^{-4}$, for a small change in strain we observe a large change in sample resistivity. Indeed, the *strain gauge factor*, defined as the fractional change in sample's resistance divided by the fractional change in sample's length, can be as high as ~ 10,000.[17] This is much larger than the typical gauge factor of standard metal-foil strain sensors which is ~ 2.

The ability to change the valley occupancy in AlAs QWs either by choosing the QW width or by applying in-plane strain makes the AlAs 2DES very flexible and exciting. It allows us to control and explore the role of a new (valley) degree of freedom in the physics of very clean 2DESs. In the following sections, we elaborate on some of the exciting problems we have studied in AlAs 2D electrons. We emphasize two important points at the outset. First, via a proper choice of the AlAs QW width *and* the applied strain, we can effectively choose to populate any combination of the three $X$, $Y$, and $Z$ valleys that we wish.[13,29] For example, by growing a QW with a width just below 5 nm, we can aim to have the $Z$ valley occupied while the $X$ and $Y$ are just empty (*i.e.*, the Fermi level lies just below the $X$ and $Y$ valley energies). Then by compressing the sample along [100], we can bring down the $X$ valley's energy and transfer electrons from $Z$ to $X$. Second, the 2DES is *qualitatively* different when the different valleys are occupied. This is particularly so when we compare the $Z$ valley with, e.g., the $X$ valley: the former is a 2DES with an *isotropic* Fermi contour and a rather small effective mass ($m_t^* = 0.20$), while the latter has an *anisotropic* contour with a much heavier mass along [100] ($m_l^* = 1.1$). The 2DES in AlAs QWs thus not only provides us with a system where the number of occupied valleys can be changed, but also lets us tune the 2DES, *in-situ*, into systems with very different, fundamental parameters.

## 3 Quantum Hall effect in AlAs quantum wells

As schematically shown in the fan diagrams of Fig. 4, there are three main energies in an AlAs 2DES when it's placed in a magnetic field. The field quantizes the allowed energies into a set of Landau levels (LLs), separated by the cyclotron energy, $E_C = \hbar eB_\perp/m_C$, where $m_C$ is the cyclotron effective mass and $B_\perp$ is the component of the magnetic field perpendicular to the sample plane. Because of the electron spin, each LL is further split into two levels, separated by the Zeeman energy, $E_Z = g^*\mu_B B$, where $B$ is the *total* magnetic field. Now in a wide AlAs QW, in the absence of in-plane strain, each of these energy levels in our 2DES should be two-fold degenerate because of the $X$ and $Y$ valley degeneracy. By applying in-plane strain, we remove this degeneracy and introduce a third energy, the splitting between the $X$ and $Y$ valleys ($E_V$) which increases linearly with strain, as illustrated in the fan diagram of Fig. 4b. On the other hand, at a fixed valley-splitting, we can tune $E_Z$ by tilting the sample by an angle θ in the magnetic field (θ is the angle between the sample normal and the field direction). This results in an increase in $E_Z$ relative to the other two energies as schematically shown in Fig. 4a fan diagram. Thanks to the relatively large values of the effective mass and g-factor of AlAs 2DES, the fan diagram of Fig. 4a can be easily achieved, *i.e.*, the LLs can be made to cross at reasonably small of values of θ compared to, e.g., GaAs 2D electrons where large θ are required because of the electron small effective mass and g-factor.

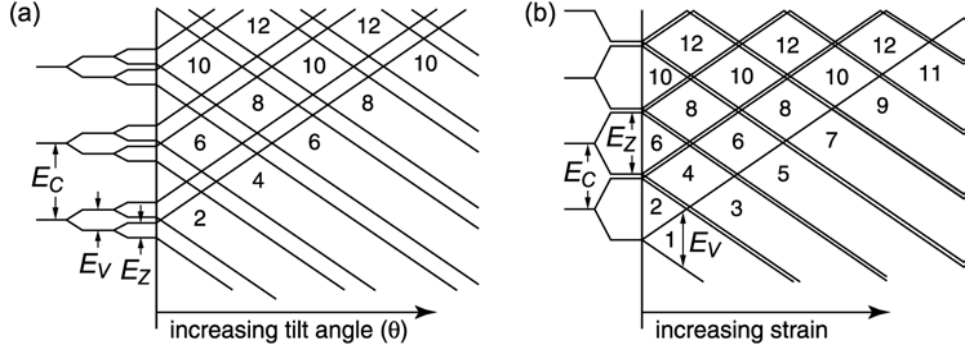

**Fig. 4** Schematic fan diagrams showing the relevant energies of a non-interacting AlAs 2DES in a magnetic field. The perpendicular component of the field determines the cyclotron energy ($E_C$), while the Zeeman energy ($E_Z$) and the valley splitting ($E_V$) can be tuned by (a) tilt-angle, or (b) strain, respectively.

### 3.1 Valley-splitting and integer QHE in wide AlAs quantum wells

Figure 5b shows a longitudinal resistivity ($\rho_{xx}$) vs. $B_\perp$ trace taken in one of our samples where the electrons are confined to an 11 nm-wide AlAs QW. The sample was glued to a piezo actuator, and the piezo was biased such that the energy splitting between the $X$ and $Y$ valleys was minimized. The resulting valley degeneracy manifests itself in the trace as strong minima in $\rho_{xx}$ are observed at low $B_\perp$ only at every fourth LL filling factor, *i.e.*, at ν = 6, 10, 14, 18, etc. In other words, there is a four-fold (near) degeneracy of the LLs (two for spin and two for valley) at low $B_\perp$. This is qualitatively consistent with what we would expect based on the fan diagram of Fig. 4b for the case of zero strain. It is in contrast to the trace of Fig. 5a which was taken on a 4.5 nm-wide AlAs QW where only the $Z$ valley is occupied: in the narrow well case, no four-fold degeneracy of the LLs is either expected or observed.

But note in Fig. 5b that at higher fields, $\rho_{xx}$ exhibits strong minima at *all* integer ν, reflecting an apparent "valley-splitting" that increases with $B_\perp$ and, at sufficiently large $B_\perp$, leads to QHE at all integer fillings. The observation of valley-splitting in a $B_\perp$ was first reported in Si-MOSFET 2DESs,[20] and remains a subject of continued research.[21] In analogy to the exchange-enhancement of the spin-splitting, the valley-splitting at high $B_\perp$ is generally believed to be a result of electron-electron interaction, although the details are not quantitatively understood.[20] In the case of AlAs 2DESs, we recently did coincidence measurements in tilted magnetic fields, and observed a valley-splitting that linearly increases with $B_\perp$.[18] This is qualitatively similar to what has been reported in Si-MOSFETs. The magnitude of the splitting, and particularly its linear dependence on $B_\perp$, however, remain puzzling: if the splitting is induced by interaction, one would expect it to increase as the inverse of the magnetic length, *i.e.*, as $B_\perp^{1/2}$.

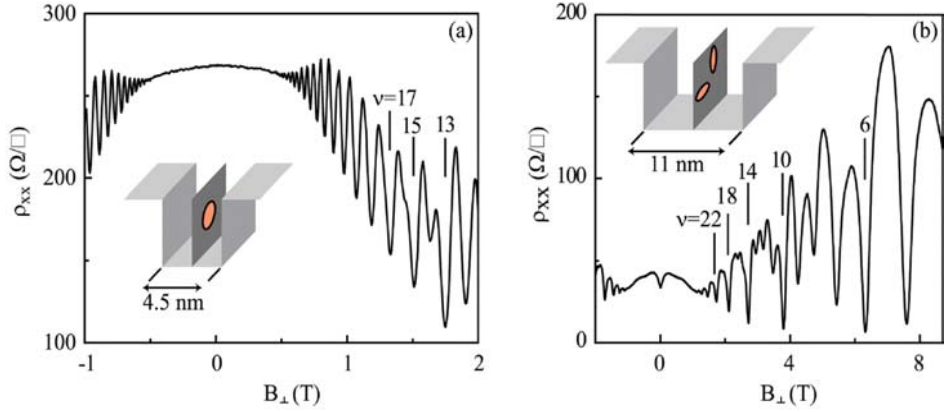

**Fig. 5** Magneto-resistance traces for 2DESs confined to (a) a narrow (4.5 nm) AlAs QW where the out-of-plane (*Z*) valley is occupied, and (b) a wide (11 nm) AlAs QW where two in-plane (*X* and *Y*) valleys are occupied. The Fermi contours of the occupied valleys are schematically shown in the insets. [After Refs. 9, 18, and 19.]

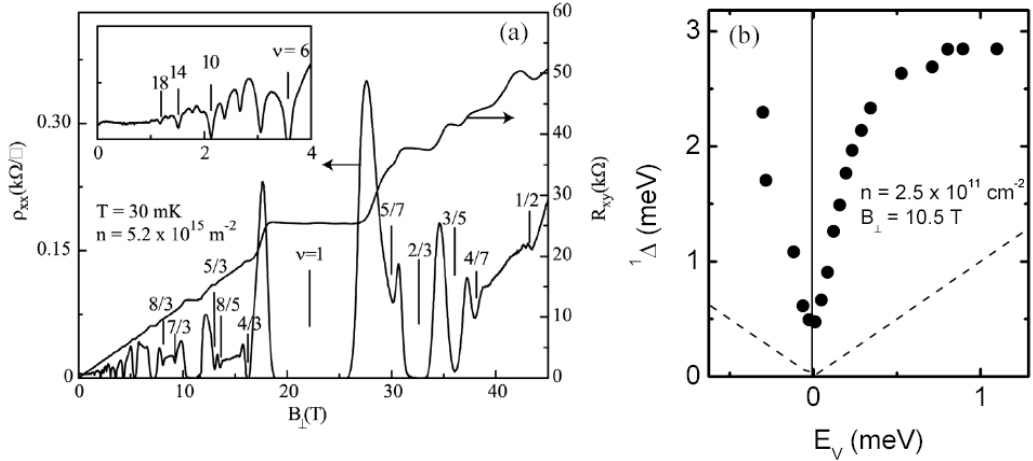

**Fig. 6** (a) Magneto-resistance traces for a 2DES confined to an 11 nm-wide AlAs QW and with the in-plane strain tuned so that the *X* and *Y* valleys are nearly equally occupied. The traces show strong QHE at ν = 1 and numerous fractional ν. The inset shows a close-up of the low field data. (b) Energy gap for the ν = 1 QHE in a similar sample vs. single-particle valley-splitting. Dashed line is the gap expected from a single-particle picture. [After Ref. 22.]

### 3.2  QHE at ν = 1: Evidence for "Valley Skyrmions"

A particularly interesting problem that is directly related to the valley-splitting is the origin of the QHE at ν = 1 in a wide AlAs QW. According to a single-particle, non-interacting picture, there should be no QHE at ν = 1 in the absence of in-plane strain (see, e.g., Fig. 4b fan diagram). There is, however, a very strong ν = 1 QHE with a sizeable energy gap even when the energy splitting between the *X* and *Y* valleys is minimized (Fig. 6).[22] The presence of this gap clearly betrays the role of interaction. Moreover, as we tune the energies of the valleys via the application of in-plane strain, the ν = 1 QHE gap rises much faster than expected from a single-particle picture (dashed line in Fig. 6b). This behavior is very reminiscent of the dependence of the ν = 1 QHE gap on the Zeeman energy in the (single-valley) GaAs 2DES where the fast rise is interpreted as evidence for spin Skyrmions (spin texture).[23] Given the similarity of the spin and valley degrees of freedom, a plausible interpretation of Fig. 6b data is that the lowest energy charged excitations at ν = 1 are "valley Skyrmions" akin to the spin Skyrmions.[22]

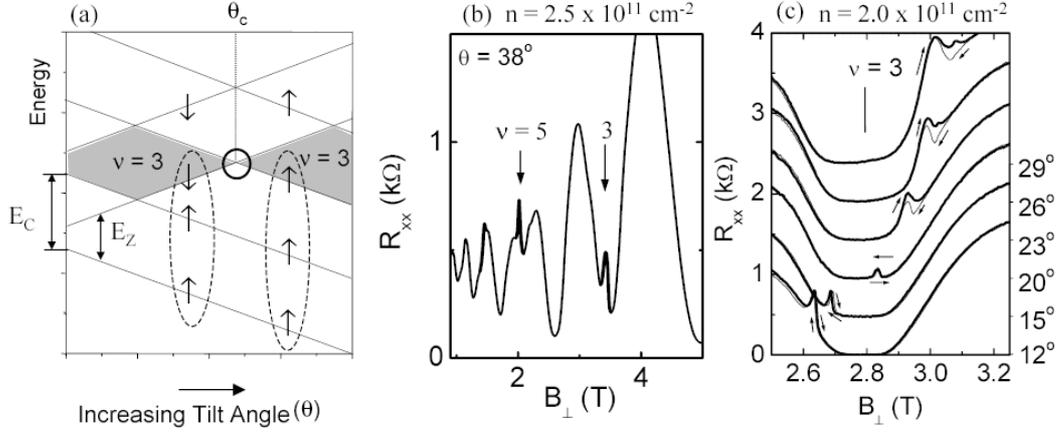

**Fig. 7** QHE ferromagnetism in a wide AlAs QW where only one in-plane valley is occupied. The schematic fan diagram in (a) shows the transition of the ν = 3 QHE state to a spin-polarized configuration at a critical tilt angle θ$_c$. Magneto-resistance traces in (b) and (c) show that the transition is signalled by a resistance "spike" which can be hysteretic when the magnetic field is swept up or down. [After Ref. 25.]

### *3.3 Quantum Hall ferromagnetsim in AlAs 2D electron systems*

Magneto-transport measurements in AlAs QWs reveal a remarkable behavior, most commonly observed[24] when we tilt the sample with respect to the direction of magnetic field. Figure 7b provides an example. The data show that near the tilt angles (coincidence angles) where the LLs of opposite spin cross at the Fermi level (e.g., near θ =38° in Fig. 7b or 23° in Fig. 7c), there are "spikes" in the resistance, signalling extra dissipation at such crossings.[25,26,8] Moreover, the position of the spike depends on θ, and the magneto-resistance trace near the spike shows hysteresis as the direction of field sweep is reversed (Fig. 7c). Such phenomena are observed in both wide[25,8] and narrow[26] AlAs QWs, and it has also been reported in other 2D systems.[27] The spikes are believed to be related to transitions between ferromagnetic QHE states and the extra disspipation one expects at domain walls that separate competing phases at the transition,[28] although the exact origin of the spikes is not yet understood. Most recently, we have studied the physics of what happens at crossings between different LLs belonging to different valleys.[29,30] We have found intriguing results, including an unexpected dependence of the magitude of the energy gaps of the QHE states that persist at LL crossings on the relative spin orientation of the crossing levels.[30]

### *3.4 Spin-dependent resistivity at transitions between QHE states*

Data of Fig. 8 show yet additional, intriguing features in magneto-transport properties of AlAs 2DESs.[26] As θ is further increased, the spike merges with the resistance maximum between the ν = 3 and 2 QHE states and, remarkably, "lifts" the resistance to values that are about an order of magnitude larger than those at small θ. The enhanced resistivity appears to be related to the orientation of the electron spin in the partially filled LL where the Fermi level resides. More specifically, the amplitude of a given resistance peak is maximal when the partially-filled LL has the same spin as the lowest LL.[26] This results in an alternating pattern for the amplitudes of the resistance maxima at transitions between the QHE states, as can be seen in Fig. 8. Which maximum is stronger and which one weaker depends on θ, as expected from a simple fan diagram like the one shown in Fig. 4a (note that for the data of Figs. 7 and 8, $E_V$ is large so that only one valley is occupied). The likely physical reason for the spin-dependent resistivity is that, because of the Pauli exclusion principle, the screening of the disorder by the electrons becomes less efficient when the spins are aligned. Such a mechanism has indeed been proposed to explain the magneto-resistance of 2D systems in a purely parallel field.[31] A question of relevance here is whether a spin-dependent screening mechanism can also explain the sharp resistance spikes that are observed deep in the QHE states (see, e.g., Figs. 7b and 7c); it appears so far that invoking an additional mechanism, such as domain wall scattering, is necessary to explain the latter. Finally, we expect that spin dependent resistivity in the QHE regime should be present in other 2D carrier systems; indeed, it has recently been reported in the Si/SiGe 2DESs.[32]

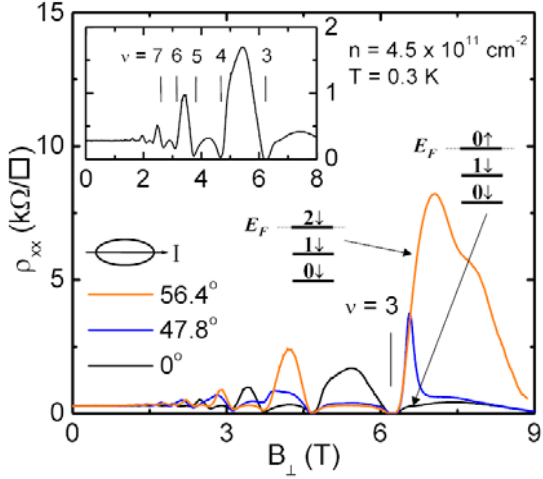

**Fig. 8** Magneto-resistance traces for a 2DES confined to a wide (11 nm) AlAs QW where only one in-plane valley is occupied. The inset shows a close-up of the θ = 0 trace. [After Ref. 26.]

### 3.5  *Fractional QHE in wide AlAs quantum wells*

Figure 6a reveals that the 2DESs in wide AlAs QWs exhibit QHE at *fractional* fillings, both in the range ν > 1, and in the extreme quantum limit (ν < 1). Note that this trace was taken at ~ 30 mK and up to 45 T in the hybrid magnet system at the Florida National High Magnetic Field Laboratory. These conditions represent the highest DC fields and lowest temperatures that are available at such high fields.  And the sample was mounted on a piezo actuator, so that we could tune the valley-splitting *in-situ*! Figure 6a $\rho_{xx}$ and $R_{xy}$ traces were taken with the piezo bias adjusted so that the splitting between the *X* and *Y* valleys was near a minimum. Nevertheless, the trace exhibits fractional QHE states at both odd- and even-numerator fillings, qualitatively similar to a standard, single-valley GaAs 2DES, and reflecting the splitting of the valleys at high $B_\perp$.[33]  (In a non-interacting picture where the valleys are degenerate and independent, we would expect fractional QHE only at *even*-numerator fillings.)  Recent results in high-mobility Si/SiGe 2DESs, where two (out-of-plane) valleys are occupied also show fractional QHE states at odd-numerator fillings although, compared to a single-valley 2DES such as one in GaAs, they are weaker than expected from the strength of their neighboring even-numerator states.[34] The data in Ref. 34 were interpreted in terms of the role of valley degeneracy and as evidence for two-flux composite Fermions. Our limited data on AlAs 2DESs indicate that indeed the strength of the fractional QHE states for ν < 1 changes with the piezo bias, and does depend on the B = 0 valley-splitting.

In the ν > 1 regime, in our samples we have observed fractional QHE at ν = 4/3, 5/3, 7/3, 8/3, 7/5, and 8/5, and developing QHE states at surprisingly large ν, such as 11/3, 13/3, and 17/3.[7] These latter states are unusual: fractional states at ν > 3 are seldom observed in GaAs 2DESs, and only in samples with extremely high quality.[35] Our studies so far indicate that the strength of the QHE states we observe at fractional ν > 1 depends sensitively on the splitting between the *X* and *Y* valleys, and demonstrate the intricate role of valley degeneracy/splitting in the many-body states of the 2DES. This is a very interesting topic which we are currently studying.[36]

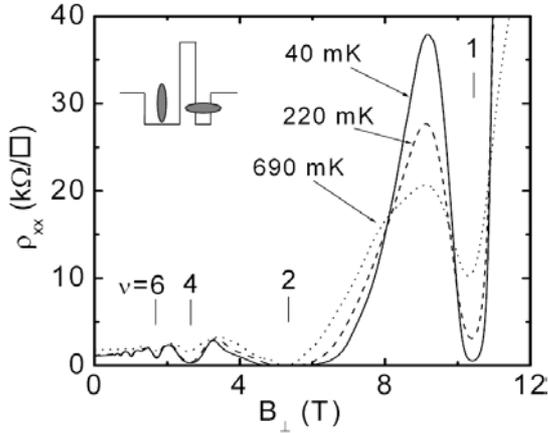

**Fig. 9** Magneto-resistance for a "hybrid" bilayer AlAs 2DES where the electrons in the two QWs occupy differently oriented valleys as indicated in the inset: the *Z* valley in the narrow QW and the *X-Y* valleys in the wide QW. [After Ref. 40.]

*3.6 Bilayer AlAs 2D electron systems*

Systems in which the carriers occupy two layers in close proximity so that the intralayer *and* interlayer interactions are both strong are among the richest in 2D physics. The system is particularly interesting and unique at the total $\nu = 1$ (layer filling ½) where it condenses into a state that possesses interlayer phase coherence and exhibits unusual properties, such as Josephson-like interlayer tunneling.[37] Recent *counterflow* measurements at $\nu = 1$, where independent contacts to each layer were made and currents of equal magnitude but opposite polaritiy were passed in the two layers,[38,39] revealed that the longitudinal *and* Hall counterflow resistances both vanish in the limit of zero temperature. This observation demonstrates the pairing of oppositely charged carriers in opposite layers (excitons), and implies that the ground state of the system is an excitonic condensate, *i.e.*, a superfluid. The AlAs 2DESs offer the possibility of realizing new types of bilayer systems. For example, consider a system where the electrons in one layer occupy the out-of-plane valley of a narrow AlAs QW while in the other they reside in in-plane valley(s) of a wide AlAs QW.[40] In such a "hybrid" bilayer system, the electrons in the two layers have very different parameters: effective mass, g-factor, and Fermi contours.[41] Moreover, because the electrons in the two layers occupy valleys located at different points of the Brillouin zone, the interlayer tunneling should be strongly suppressed even if the layers are very closely spaced. While the close spacing is needed for the stability of the interlayer-interaction-induced states such as the bilayer QHE at $\nu = 1$, the reduced tunneling is particularly desirable for experiments (like drag or counterflow) where independent contacts to the two layer are to be made.

Recently, we succeeded in making such a "hybrid" bilayer system, and in it we even observed a phase-coherent, bilayer $\nu = 1$ QHE state![40] The data, shown in Fig. 9, were taken in a parallel-flow geometry where the current was passed in the same direction through both layers. The experiments additionally revealed that the strength of the $\nu = 1$ QHE is independent of the parallel field component, implying that there is very little tunneling between the two layers even though the barrier separating them is only 2.8 nm thick. (For comparison, in bilayer systems in GaAs double QWs where tunneling was sufficiently small so that independent contacts could be achieved, the barrier was at least about 10 nm thick).[38,39] Making independent contacts to 2DESs in two AlAs layers, however, is challenging because illumination and back/front gate biasing are often needed to populate the two QWs[11] but it should be possible in principle. Closely spaced bilayer systems in which the layers have different masses might also host a novel superconducting state at zero magnetic field.[42] Bilyer AlAs 2DESs are thus clear targets for future research.

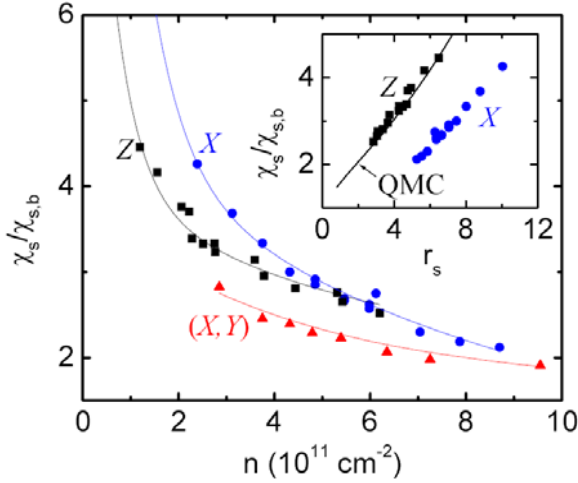

**Fig. 10** Spin susceptibility $\chi_s$, normalized to its band value $\chi_{s,b}$, plotted as a function of 2DES density for AlAs QWs with different valley occupations. All the data points were measured in tilted magnetic fields via the coincidence method. Lines through the data points are guides to the eye. The inset shows the normalized susceptibility, for the cases where only the *X* or *Z* valley is occupied, vs. the interaction parameter $r_s$; the results of quantum Monte Carlo (QMC) calculations, shown by the solid curve, quantitatively agree with the *Z* valley data. [After Refs. 9 and 43.]

## 4 Spin susceptibility of AlAs 2D electrons

A highlight of our work has been our study of the spin susceptibility $\chi_s$ in AlAs 2DESs. The dependence of $\chi_s$ on electron density, *n*, especially at very low *n* where the interaction dominates, has been a topic of considerable interest recently, partly because of its possible connection to the metal-insulator transition problem in 2D systems.[14] We measured $\chi_s$ in both narrow[9] and wide[43] AlAs QWs as a function of *n* (Fig. 10). The narrow QW 2DES should be closest to an "ideal" 2D system (except for disorder of course): (*i*) only one valley is occupied, (*ii*) the in-plane Fermi contour is circular and therefore isotropic, (*iii*) it is very narrow (QW width $\cong$ 4.5 nm) so that finite layer thickness corrections[44] should be negligible, and (*iv*) the dilute limit can be reached relatively easily because the electron effective mass for this system ($m_t^* = 0.20$) is reasonably large compared to, e.g., $m^* = 0.067$ for GaAs 2D electrons, so that large values of $r_s$, the average inter-electron distance measured in units of the effective Bohr radius, can be attained. We found that at a given $r_s$ this 2DES exhibits the largest enhancement of $\chi_s$ among all the 2DESs and, even more remarkably, the enhancement is in excellent agreement with the results of quantum Monte Carlo calculations[45] without any adjustable parameters (Fig. 10 inset).[9] This finding appears to reflect the fact that the 2DES in a narrow AlAs QW is indeed closest to ideal.[46]

In another set of experiments, we studied $\chi_s$ in wider AlAs QWs where the *X* and/or *Y* valleys are occupied (Fig. 10). We applied in-plane stress to transfer electrons from the *Y* valley to *X* and measured $\chi_s$ as a function of valley occupation.[43] We found that, at a given *n*, $\chi_s$ is *larger* when the electrons are all transferred into one valley (*X* in Fig. 10 data). This observation counters the common assumption that a two-valley 2DES is effectively more dilute than a single-valley system because of its smaller Fermi energy! Our work has already stimulated theoretical work, aiming to explain this surprising observation based on subtle electron-electron interaction.[47]

Another intriguing feature of the data is that, when plotted as a function of $r_s$ (Fig 10 inset), $\chi_s$ is smaller for the *X* valley compared for the *Z* valley. This may be partly due to the larger width ($\cong$ 11 nm) of the QW in the case where the *X* valley is occupied, but finite layer thickness is unlikely to fully account for the discrepancy. It is more likely that the culprit is the anisotropic Fermi contour (and therefore the anisotropic in-plane effective mass) of the 2DES when the electrons occupy only the *X* valley. If so, this would be a clear indication that anisotropy plays a crucial role in electron-electron interaction.

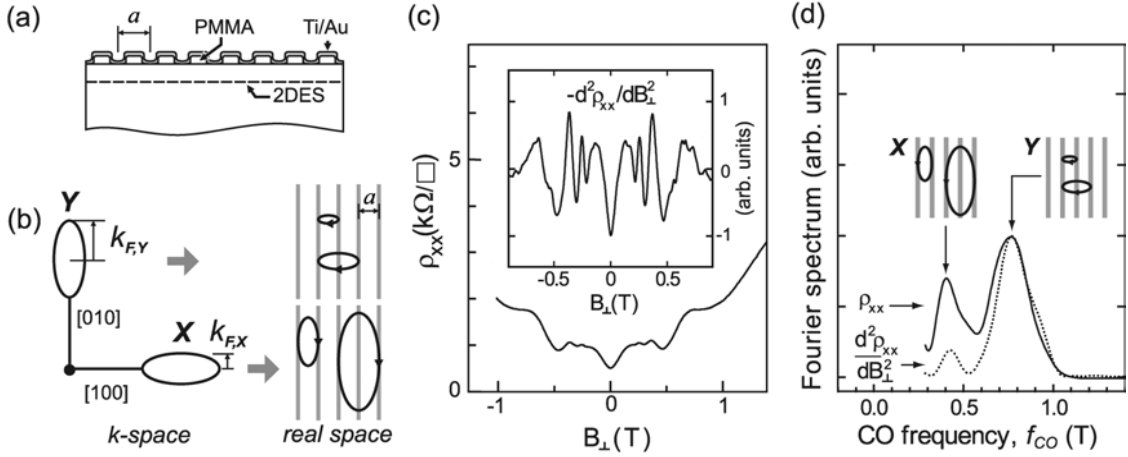

**Fig. 11** Summary of commensurability oscillations (COs) in a wide AlAs QW where the *X* and *Y* valleys are occupied. (a) Sample cross-section showing how the 2DES density is modulated with a period of *a*. (b) Because of their different orientations, the conditions for the COs of *X* and *Y* valley electrons are met at different values of $B_\perp$, leading to two different frequencies of the COs. (c) Examples of COs in resistivity and its second derivative (inset), and (d) Fourier spectra of the of the traces in (b) showing the two frequencies of the COs. [After Ref. 50.]

## 5  Ballistic transport of AlAs 2D electrons

The rather high mobility of the 2DES in wide AlAs QWs allows us to explore ballistic electron transport in this system. We have observed, e.g., quantized steps in the resistance of a point-contact device made in an AlAs 2DES.[48,49] The data reveal interesting features, including a well-developed "0.7 structure" which may reflect the role of strong interaction in the system.[48] As summarized in Fig. 11, we have also measured the commensurability oscillations (COs) by patterning a surface-gate structure (Fig. 11a) where the 2D electrons move in a potential which is modulated along [100].[50] As is well known from the measurements of COs in standard GaAs 2DESs,[51] the resistance in such a structure oscillates as a function of $B_\perp$. The frequency of the oscillations is proportional to the electron Fermi wave vector along the modulation direction. Now, in the AlAs 2DES, the *X* and *Y* valleys each have an elliptical orbit and the major axes of these orbits are orthogonal to each other (Fig. 11b). One would then expect two sets of oscillations with frequencies $f_X$ and $f_Y$ whose ratio squared gives the ratio of the longitudinal and transverse effective masses: $(m_l^*/m_t^*) = (f_Y/f_X)^2$ (see Ref. 50 for details). This is indeed observed in the experiments (Figs. 12b,c), yielding a value of $5.2 \pm 0.5$ for the ratio of the effective masses. Note that such a measurement uniquely probes the mass ratio and is complementary to a cyclotron resonance measurement which determines the cyclotron mass $m_{CR} = (m_l^* m_t^*)^{1/2}$. Using the measured $m_{CR} = 0.46$ in wide AlAs QWs,[5] we can use the above ratio to deduce $m_l^* = 1.1 \pm 0.1$ and $m_t^* = 0.20 \pm 0.02$ for AlAs.[50]

## 6  Concluding remarks

We have described the realization and physics of 2DESs confined to AlAs QWs. Compared to the standard 2DESs in GaAs, the AlAs 2D electrons possess a much larger (and anisotropic) effective mass, a much larger effective g-factor, and occupy conduction band valleys whose populations and orientations can be controlled. These characteristics add new twists to the fundamental problem of many-body effects in confined, low-disorder electronic systems. One theme that has emerged from our study so far is the remarkable similarity of the valley and spin degrees of freedom. This can be seen, e.g., from the enhanced valley-splitting in a $B_\perp$ which is akin to exchange-enhanced spin-splitting, or in the Skyrmion problem near ν = 1 (Fig. 6b). Given that valley and spin are SU(2) symmetric degrees of freedom, this similarity is not unexpected. Note also that symmetry-breaking strain plays the same role for the valley degree of freedom that (total) magnetic field plays for spin. Our most recent work provides perhaps the clearest demonstration of the near equivalence of spin and valley degrees of freedom. In a 2DES in a wide AlAs QW we measured the "valley-susceptibility," the change of valley population in response to applied in-plane strain, and found that it increases dramatically as the density is reduced.[52] The increase is quantitatively similar to the enhancement of the spin-susceptibility seen in Fig. 10. The general analogy between the valley and spin degrees of freedom implies the potential use of valleys, in lieu of or in addition to spin, for applications such as spintronics and quantum computing.

**Acknowledgements**   Our work was supported by the NSF, DOE, and ARO. We also thank N.C. Bishop, T. Gokmen, K. Karrai, S. Manus, S. Misra, M. Padmanabhan, and S.J. Papadakis for their contributions to this work.